\documentclass[runningheads]{llncs}
\usepackage{orcidlink}
\usepackage[T1]{fontenc}
\usepackage{cite}
\usepackage{amsmath,amssymb,amsfonts}
\usepackage{algorithmic}
\usepackage{graphicx}
\usepackage{textcomp}
\usepackage{xcolor}
\usepackage{hyperref}
\usepackage{academicons} 
\usepackage{graphicx}
\begin{document}
\title{A large language model system for the field of chemical engineering technology
}

\titlerunning{A LLM system for the field of chemical engineering technology}
%
%
\author{
  Heng Zhang\inst{1,}$^a$ \orcidlink{0009-0009-3871-861X} \and
  Jibin Zhou\inst{3,}$^a$ \orcidlink{0000-0002-9269-119X} \and
  Feiyang Xu\inst{4,5} \orcidlink{0000-0002-9627-7930} \and
  Jian Cui\inst{4,5} \orcidlink{0009-0009-3968-8061} \and
  Yi Li\inst{4,5} \orcidlink{0009-0004-0261-9197} \and
  Fan Yang\inst{4,5} \orcidlink{0009-0002-6401-8333} \and
  Hao Wang\inst{1,5,}$^*$ \orcidlink{0000-0001-9921-2078} \and
  Xin Li\inst{2,4,5,6,}$^*$ \orcidlink{0000-0002-7333-5114} \and
  Mao Ye\inst{3,}$^*$ \orcidlink{0000-0002-7078-2402}
}
\authorrunning{H. Zhang et al.}
%
\institute{School of Computer Science and Technology, University of Science and Technology of China, Hefei, China \\
\email{\{wanghao3, leexin\}@ustc.edu.cn} \and
School of Information Science and Technology, University of Science and Technology of China, Hefei, China \and
National Engineering Research Center of Lower-Carbon Catalysis Technology, Dalian Institute of Chemical Physics, Chinese Academy of Sciences, Dalian, China \\ 
\email{maoye@dicp.ac.cn} \and
Artiffcial Intelligence Research Institute, iFLYTEK Co., Ltd., Hefei, China \and
State Key Laboratory of Cognitive Intelligence, Hefei, China \and
IFLYTEK (INTERNATIONAL) LIMITED, Hong Kong, China \\
$^a$ These authors contributed equally to this work.  
}
\maketitle              
\begin{abstract}
The development of chemical engineering technology is a multi-stage process that encompasses laboratory research, scaling up, and industrial deployment. This process demands interdisciplinary collaboration and typically incurs significant time and economic costs. To tackle these challenges, we have developed a system based on ChemELLM in this work. This system enables users to interact freely with the chemical engineering model, establishing a new paradigm for AI-driven innovation and accelerating technological advancements in the chemical sector.If you would like to experience our system, please visit our official website at: https://chemindustry.iflytek.com/chat.

\keywords{chemical engineering  \and large language model.}
\end{abstract}
\section{Introduction}
The development of chemical engineering technology is a multi-stage progression, spanning laboratory research, scaling-up, and the advancement of foundational engineering, before ultimately achieving industrial deployment \cite{r1,r2}. This intricate process demands collaboration among experts from diverse disciplinary backgrounds—including chemistry, physics, mathematics, electronics, and computer engineering—while balancing economic feasibility with resolving technical challenges. Nevertheless, interdisciplinary collaboration remains hindered by disciplinary boundaries, presenting obstacles to ensuring consistency throughout the development of chemical processes \cite{r3}.

Recently, emerging strategies such as data-driven AI technologies have gained increasing recognition for their potential to streamline development processes and enhance efficiency \cite{r4, r5}. Particularly, the advent of large pre-trained models, trained on vast interdisciplinary corpora, offers greater possibilities for optimizing scientific workflows\cite{r6, r7}. However, current chemical LLMs focus primarily on molecular-scale tasks and are ineffective in addressing challenges in the system engineering domain. There remain significant limitations in solving core chemical engineering issues, such as process simulation, equipment design, and industrial-scale optimization, thus developing large models for chemical engineering has become an urgent task.

In the development of domain-specific large language models (LLMs), systematically evaluating their ability to understand and apply domain knowledge is crucial. Within the broader field of chemistry, several benchmarks including ChemLLMBench \cite{r17}, SciBench \cite{r31}, and ChemEval \cite{r32} have been established to rigorously assess LLMs’ performance in chemical concept explanation, logical reasoning, and complex problem-solving. However, these benchmarks have limitations in evaluating LLM capabilities specific to chemical engineering, particularly when it comes to addressing the core competencies required for tackling industrial-scale challenges. Thus, there is an urgent need to develop specialized benchmarks for systematically assessing the practical application capabilities of LLMs in chemical engineering. This will facilitate their effective deployment in key technical areas such as catalyst design, fluid dynamics simulation, process optimization, and equipment selection.

\begin{figure*}
    \centering
    \includegraphics[width=1\linewidth]{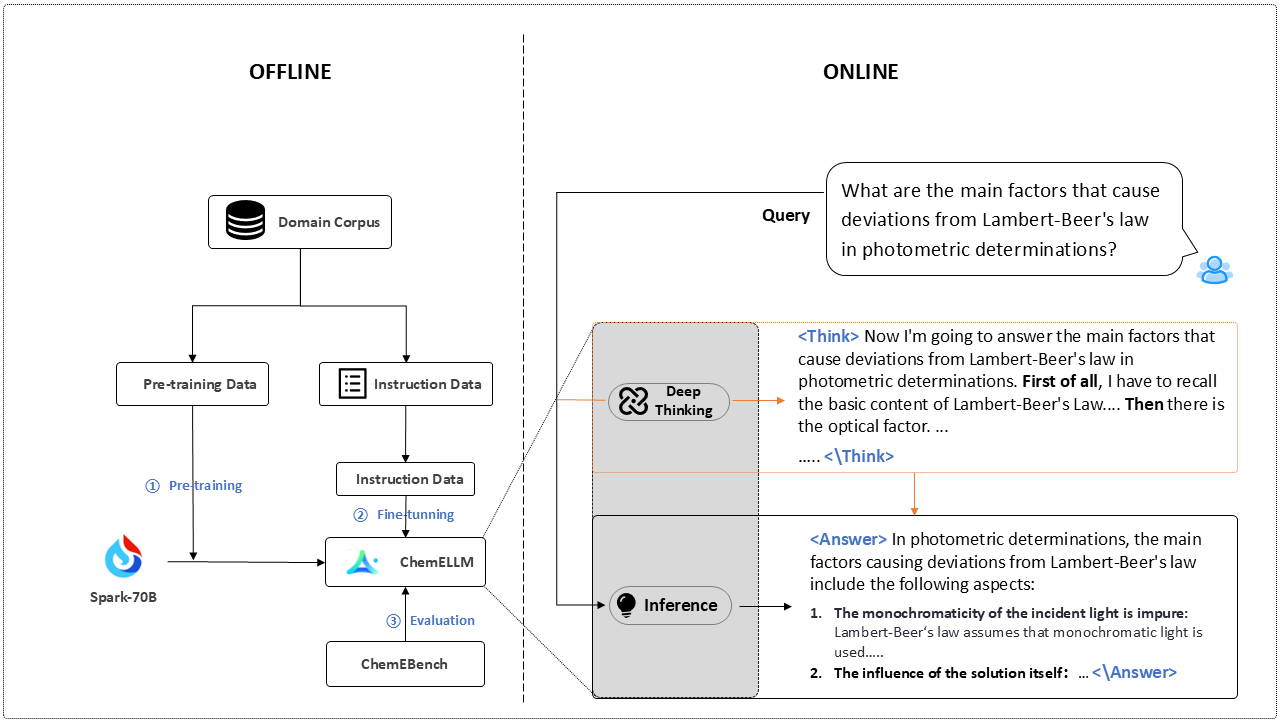}
    \caption{System architecture diagram}
    \label{fig1}
\end{figure*}

This paper builds upon the work of the first domain-specific LLM designed for chemical engineering applications (ChemELLM) \cite{zhou2025lab}, integrating a large chemical model platform. ChemELLM uses the Spark-70B base model, leveraging chemData (a meticulously curated high-quality chemical engineering corpus) for domain-adaptive pre-training and instruction fine-tuning.

\section{System Framework}
\begin{figure}
    \centering
    \includegraphics[width=1\linewidth]{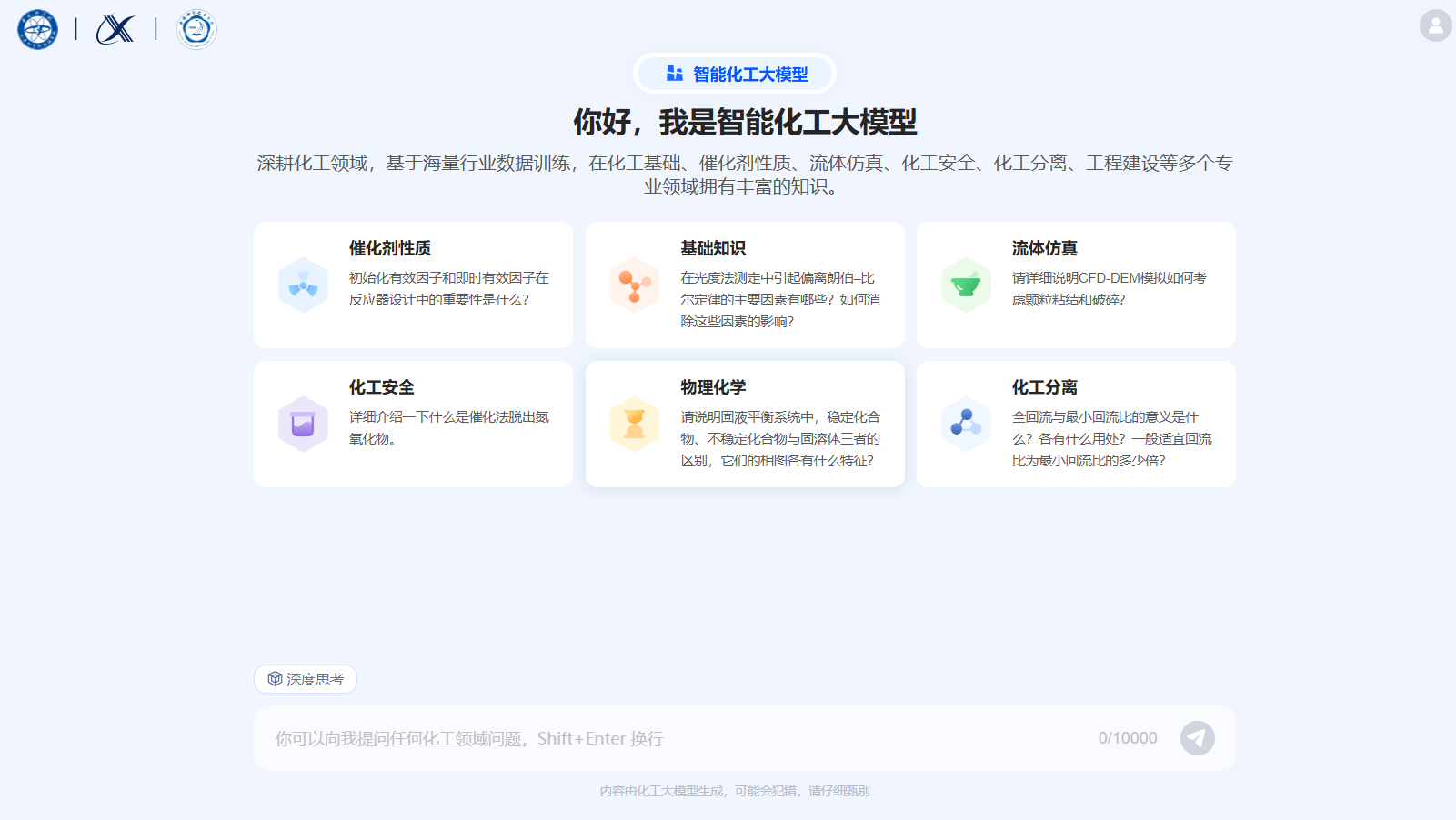}
    \caption{System interface diagram}
    \label{fig2}
    \vspace{-10pt}
\end{figure}

In this section, we present the framework of the large chemical model, as illustrated in Fig. \ref{fig1}. The entire system is divided into two components: an online component and an offline component. The offline component primarily describes the training process of the model using interdisciplinary corpora, whereas the online component outlines how the chemical model system processes user input data.
As depicted in Fig. \ref{fig2}, users can seamlessly access the system via the intuitive interactive interface.
\subsection{Online Part}
\subsubsection{User Input}
In the system described in this paper, users can input chemical engineering queries in the form of text or voice, covering multiple subfields such as catalyst properties, fundamental knowledge, fluid simulation, chemical engineering safety, physical chemistry, and chemical separation. The system supports input of up to 10,000 characters.

\subsubsection{Question Inference}
Question inference, as the core component of ChemELLM in addressing queries in the chemical domain, consists of two closely linked and mutually supportive parts: ChemELLM Inference and Deep Thinking. This architectural design aims to balance the efficiency of the model's basic operations with the rigor in handling complex problems. Among them, ChemELLM Inference forms the basic operational framework for question response, responsible for completing the technical process from original query to initial answer generation; while Deep Thinking, as an advanced processing mechanism, endows the inference results with higher accuracy, coherence, and interpretability through logical decomposition and self-iteration. Together, they form a complete "basic processing - in-depth optimization" inference loop, ensuring that the model can not only quickly respond to simple queries but also properly handle complex problems involving multi-dimensional knowledge associations.

ChemELLM Inference: After a user inputs a query, the large chemical model first preprocesses the query, including tokenization, labeling, and vector representation, converting it into a vector form that the chemical model can understand. Then, through multi-layer neural networks and techniques such as attention mechanisms, it deeply processes and analyzes the semantic and grammatical information of the input. Finally, in the decoding phase, based on the output of the semantic representation by the encoder, it generates the corresponding answer, calculates the probability distribution of the output tokens using functions like Softmax, and selects the appropriate token sequence as the final output result to present to the user.

Deep Thinking: In contrast to ordinary inference, when confronted with a user’s query, ChemELLM first decomposes the question into multiple interconnected aspects, constructing a logical chain of thought. It systematically identifies all potential sub-questions that the original query might entail, breaks down the complex problem into manageable smaller questions, and addresses each with targeted analysis. Ultimately, the model meticulously reviews the handling of each sub-question and rigorously evaluates the coherence and accuracy of the final integrated outcome.During the problem-solving process, ChemELLM conducts continuous self-assessment, engaging in explicit self-affirmation or critical self-denial to determine whether sub-questions require reprocessing—repeating the cycle until it achieves a internally consistent and convincing conclusion. Throughout this iterative inference process, the model transparently displays its detailed reasoning steps, enabling users to clearly trace the entire chain of logic from initial query to final answer.

\subsubsection{Result Return}

The output content, shown on the right side of Fig. \ref{fig1}, varies based on whether the deep thinking feature is enabled. When activated, the system returns both the step-by-step reasoning process—with clear logical progression to trace the model’s cognitive path—and the corresponding answer, enhancing transparency and aiding users in understanding the derivation. 

If the feature is not activated, the system directly outputs the inference result, optimized for efficiency to meet needs for quick answers in time-sensitive or straightforward scenarios. This dual-mode design balances comprehensiveness and efficiency, adapting to diverse user needs in chemical engineering applications, from in-depth research to practical operations.

\subsubsection{Example}
In this paper, we present the answers provided by the large model to two questions as examples. 

\begin{figure}
    \centering
    \includegraphics[width=1\linewidth]{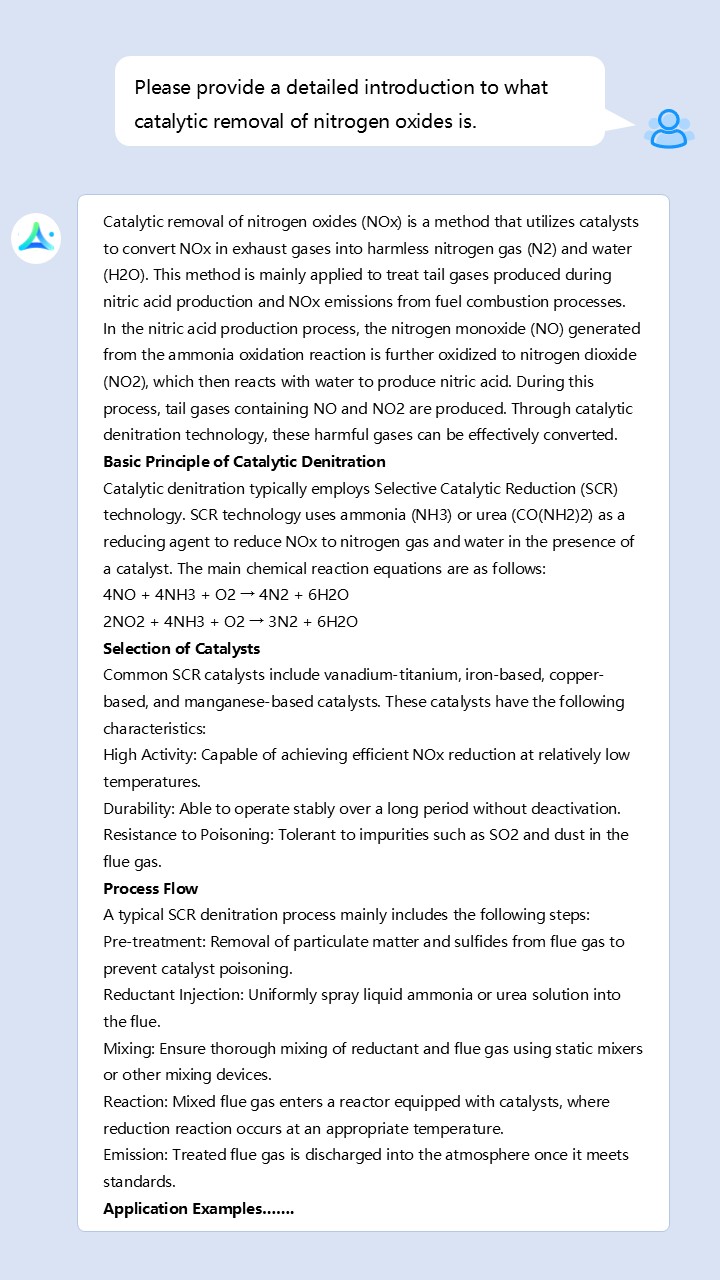}
    \caption{ChemELLM’s response to the question}
    \label{fig3}
\end{figure}

In Fig. \ref{fig3}, when a user poses a question about catalytic removal of nitrogen oxides in chemical engineering technology, the large model responds comprehensively. It covers the method's principle of using catalysts to convert NOx, its application in treating nitric acid production tail gases and fuel combustion emissions, and also mentions key aspects like catalyst selection and process flow, and would provide application examples.

\begin{figure}
    \centering
    \includegraphics[width=1\linewidth]{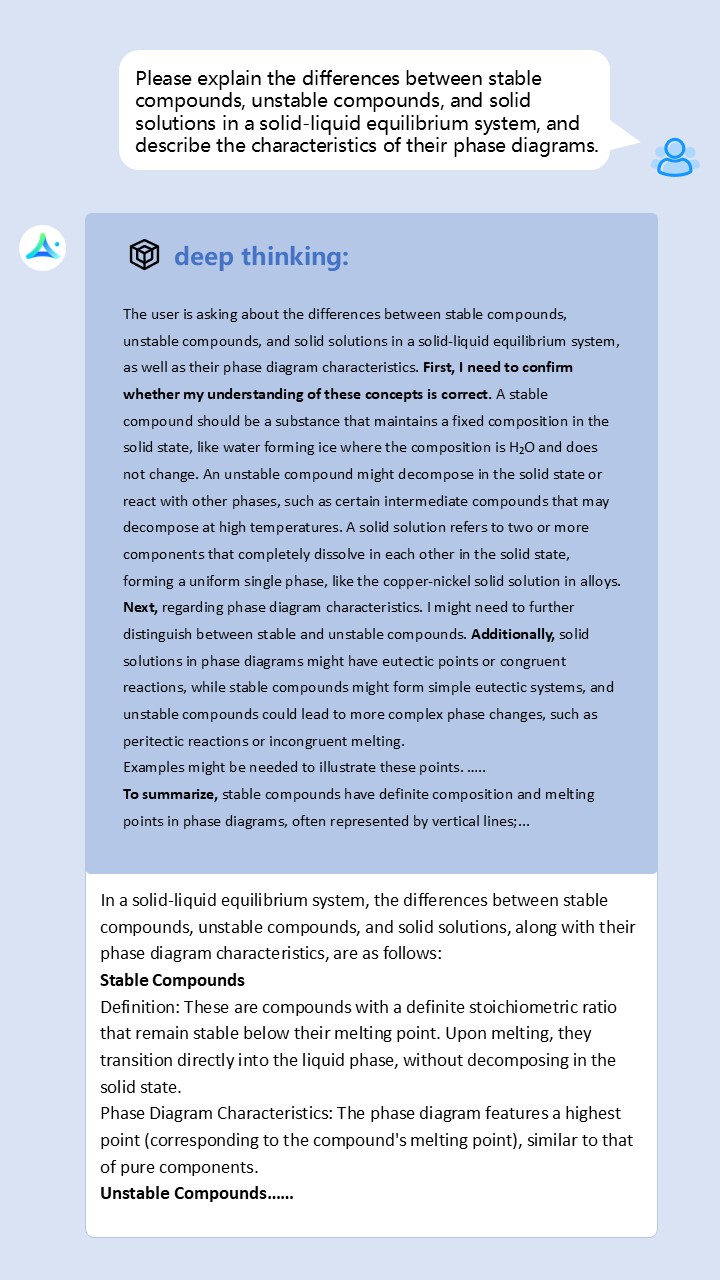}
    \caption{Example of Using Deep Thinking}
    \label{fig4}
\end{figure}

In Fig. \ref{fig4}, users are afforded the flexibility to decide whether to engage in deep thinking according to their specific needs. When the deep thinking option is chosen, the large model showcases a comprehensive and meticulous detailed reasoning process. This process is not merely a simple listing of steps but involves a thorough exploration of the problem from multiple angles. For instance, when addressing the differences between stable compounds, unstable compounds, and solid solutions in a solid - liquid equilibrium system, the model first starts by confirming its understanding of the core concepts. It carefully defines each term, like stable compounds maintaining a fixed composition in the solid state, unstable compounds potentially decomposing or reacting in the solid state, and solid solutions being a uniform single phase formed by complete dissolution of components in the solid state. Then, it delves into the phase diagram characteristics, distinguishing between stable and unstable compounds, and considering how solid solutions manifest in phase diagrams, such as through eutectic points or congruent reactions. Throughout this, the large model continuously reflects on its reasoning steps. It checks if the definitions are accurate, if the connections between concepts and phase diagram features are logical, and if there are any gaps in the explanation. By doing so, it ensures that each part of the reasoning is sound, which in turn enhances the accuracy and reliability of the final answers. This deep thinking process serves as a valuable tool for users, as it not only provides the result but also allows them to follow along the model's cognitive journey, thereby gaining a much deeper and more insightful understanding of the intricate issue in the field of chemical engineering thermodynamics.

\subsection{Offline Part}
This section's work is primarily based on the previous work by \cite{zhou2025lab}. The prior work developed a vertical large language model for the chemical engineering field, ChemELLM, and constructed the first multidimensional evaluation benchmark system for chemical engineering, ChemEBench. Experiments have demonstrated that ChemELLM surpasses mainstream large language models in key chemical engineering tasks. The offline part is introduced in three main sections, including domain pre-training, supervised fine-tuning, and ChemEBench.
\subsubsection{Domain Pre-training}

Datasets used to train general large language models (LLMs) typically cover a wide range of topics but remain relatively shallow in any specific domain. As a result, while these models have successfully acquired strong natural language understanding and reasoning capabilities, they often reveal limitations when confronted with tasks demanding profound professional knowledge. To address the shortcomings of general language models in specialized fields, we conducted domain-specific pre-training on the foundational language model Spark-70B, utilizing a comprehensive chemical engineering corpus comprising 19 billion tokens. This approach enables ChemELLM to retain the inherent foundational capabilities of Spark-70B while additionally acquiring domain-specific knowledge, thereby effectively overcoming the professional bottlenecks of general models.

\subsubsection{Supervised Fine-tuning}
During the supervised fine-tuning (SFT) phase, our goal is to align the large chemical model with specific language patterns and terminology prevalent in chemical engineering. To achieve this, we employed 2.75 million high-quality data points, totaling 1 billion tokens for fine-tuning. The optimization process utilized the Adam optimizer with an initial learning rate of \(1 \times 10^{-5}\),
and a cosine decay strategy to adjust the learning rate during fine-tuning. The training process was executed on 128 Huawei Ascend 910B GPUs over three epochs, balancing computational efficiency and model integration. At this stage, high-quality SFT data enhances the chemical model's understanding of chemical engineering tasks, thereby facilitating the resolution of knowledge-based problems in the chemical engineering field.

\subsubsection{ChemEBench}
The ChemEBench benchmark comprises three progressive stages aimed at comprehensively evaluating the capability of LLMs in this specialized domain. The three stages are basic knowledge level, advanced knowledge level, and professional skill level. The basic knowledge level assesses proficiency in understanding fundamental concepts in chemical engineering; the advanced knowledge level aims to demonstrate the model's advanced professional level, surpassing basic concepts and extending into more complex areas of chemical engineering; the professional skill level evaluates the model's ability to handle complex tasks, including problem-solving in real-world scenarios and the practical application of chemical engineering knowledge. By integrating these three levels, ChemEBench provides a structured and comprehensive evaluation framework. This framework not only ensures the model has a solid foundation in chemical engineering basics but also confirms its capability to exhibit advanced reasoning and practical application skills required to handle professional-grade tasks within the discipline.

\section{Demonstration}
We will begin the demonstration by explaining the framework of the large chemical model and introducing its key features. Following this, participants will be invited to interact with the system firsthand.
The demonstration process consists of two primary steps. First, users input a question related to chemical engineering into the system and select whether to enable deep thinking functionality. Second, the large model performs inference based on the user’s input, with the chemical model’s output results then displayed on the webpage.

\section{Conclusion and Future works}
In this work, we have introduced the large chemical model system—a domain-specific platform developed specifically for chemical engineering applications.

For future work, we aim to enhance the causal reasoning and multi-modal capabilities of the large chemical model system, while integrating online search and knowledge augmentation technologies. These improvements will enable the system to evolve into a more robust and versatile tool, thereby accelerating innovation in both research and industrial applications within chemical engineering. Additionally, the official release of the system is forthcoming.

\section*{Acknowledgment}

The authors gratefully acknowledge the financial support provided by the Strategic Priority Research Program of the Chinese Academy of Sciences (Grant No. XDA0490200), the Science and Technology Innovation Project of Anhui Province (Grant No. 202423k09020010), and the National Natural Science Foundation of China (Grant No. 62441239), which jointly enabled the experimental investigations and engineering analyses reported in this work. We sincerely thank the project team for their valuable resources and technical support, which significantly contributed to the completion of this research.

\bibliographystyle{splncs04}   
\bibliography{ref}   

\end{document}